\pdfoutput = 1
\documentclass[10pt]{article}
\usepackage{spconf,amsmath,graphicx}
\usepackage[table,xcdraw]{xcolor}
\usepackage[super]{nth}

\title{Speech Recognition using EEG signals recorded using dry electrodes}
\name{Gautam Krishna$^{\star}$ \qquad Co Tran$^{\star}$ \qquad Mason Carnahan$^{\star}$ \qquad Morgan M Hagood$^{\dagger}$ \qquad Ahmed H Tewfik$^{\star}$}
\address{$^{\star}$Brain Machine Interface Lab, The University
of Texas at Austin \\
$^{\dagger}$ College of Natural Sciences, UT Austin\\}
\begin{document}
%
\maketitle
\begin{abstract}
In this paper we demonstrate speech recognition using electroencephalography (EEG) signals obtained using dry electrodes on a limited English vocabulary consisting of three vowels and one word using a deep learning model. We demonstrate a test accuracy of \textbf{79.07 \%} on a subset vocabulary consisting of two English vowels. Our results demonstrate the feasibility of using EEG signals recorded using dry electrodes for performing the task of speech recognition.  
\end{abstract}
\section{Introduction}
\label{sec:intro}

Speech recognition using non-invasive neural electroencephalography (EEG) signals is an emerging area of research. In recent years, works from different research groups have demonstrated the feasibility of using EEG features for performing isolated classification based speech recognition with high accuracy. Some of these works are described in references \cite{krishna2019speech,yoshimura2016decoding,kumar2018envisioned,sun2016neural,kim2014eeg,rosinova2017voice}. Even though EEG signals offer poor signal to noise ratio (SNR) and spatial resolution compared to invasive electrophysiological monitoring techniques like electrocorticography (ECoG) and local field potentials, the non-invasive nature of EEG makes it safe, easy to deploy and study. The EEG signals can be easily recorded by placing EEG sensors on the scalp of the subject and the signals offer high temporal resolution. 
The prior published studies on speech recognition using EEG, used EEG signals recorded using wet EEG electrodes. Even though EEG signals recorded using wet electrodes demonstrate relatively higher SNR compared to EEG recordings obtained using dry electrodes, the time required to set up a wet EEG based recording system is significantly more compared to a dry EEG system.  
Before starting a wet EEG recording session a conductive gel need to be applied on the scalp of the subjects. After the EEG recording is completed, the subjects then need to remove the gel from their scalp. In this paper we demonstrate speech recognition using EEG signals recorded using dry EEG electrodes. Dry EEG electrodes doesn't require application of conductive gel on the scalp of the subject and moreover for our experiments we used a dry EEG amplifier and wireless transmitter integrated into an easy-to-use, self-contained headset making it extremely convenient for the subjects to wear. 

Speech recognition using EEG signals might help people with speaking disabilities to restore their normal speech. Current state-of-the-art automatic speech recognition (ASR) system used in virtual personal assistants like Siri, Alexa, Bixby etc can recognize only acoustic features and this limits technology accessibility for people with speaking disabilities. Thus speech recognition using EEG can improve technology accessibility. 
References \cite{krishna20,krishna2020synthesis} studied the problems of continuous speech recognition and speech synthesis using EEG signals recorded using wet EEG electrodes. In this paper we limit our focus to the problem of isolated speech recognition. 

In this paper we propose deep learning models inspired from \cite{krishna2019speech,krishna2019improving} to perform isolated speech recognition using EEG signals recorded using dry electrodes. We were able to achieve a test accuracy as high as \textbf{79.07 \%}. We demonstrate our results on a limited English vocabulary consisting of three vowels and one word. 
Our results demonstrate the feasibility of using EEG signals recorded using dry electrodes for performing the task of isolated speech recognition using deep learning models. The experiments and models proposed in this paper can be extended to study the problems of continuous speech recognition, speech synthesis using EEG signals recorded using dry electrodes.

\section{Speech Recognition Model}
\label{sec:format}
The architecture of the speech recognition model used in this work is explained in Figure 1. The model takes dry EEG features as input and predict the label of the text. The model consists of a single layer of gated recurrent unit (GRU) \cite{chung2014empirical} with 256 hidden units connected to a dropout regularization \cite{srivastava2014dropout} with dropout rate 0.1 followed by another layer of GRU with 128 hidden units followed by a single layer of temporal convolutional network (TCN) \cite{bai2018empirical} with 32 filters. The last time step output of the TCN layer is passed to a dense layer with softmax activation function. The number of units in the dense layer can be 2 or 3 or 4 depending on the number of labels used in that experiment. The labels were one hot vector encoded. 

The model was trained for 1000 epochs with batch size 100. We used categorical cross entropy as the loss function and adam \cite{kingma2014adam} as the optimizer. Motivated by the results demonstrated by the authors in \cite{krishna2019improving}, we initialized the weights of the GRU layers of the speech recognition model using GRU layer weights derived from the regression model used to predict acoustic features from dry EEG features. The GRU layers were frozen and set to non-trainable in the speech recognition model. The intuition here is that the two GRU layers will learn the mapping from EEG to acoustic features and the trainable TCN layer will learn the mapping of these acoustic features to text. In \cite{krishna2019improving} authors used similar idea for the task of continuous speech recognition using EEG signals. 

The architecture of the regression model is described in Figure 2. The model consists of two layers of GRU with 256, 128 hidden units respectively with a dropout regularization of dropout rate 0.1 applied between the GRU layers, followed by a time distributed dense layer with linear activation function. The number of hidden units in the time distributed dense layer depends on the dimension of the target acoustic features. The model was trained to predict either mel-frequency cepstral coefficients (MFCCs) of dimension 13 or gammatone frequency cepstral coefficients (GFCCs) of dimension 13 or concatenation of GFCC and MFCC of dimension 26. In \cite{krishna2019improving} authors trained their regression model to predict MFCC and articulatory features. In this work we didn't include articulatory features as we were working with only clean data set whereas in \cite{krishna2019improving} authors used data sets recorded in presence and absence of background noise, hence articulatory features were helpful to them in providing noise robustness. 

The regression model was trained for 2000 epochs with batch size 100. We used mean squared error as the loss function and adam as the optimizer. The objective of training the regression model was to derive weights to initialize the recurrent layers in the speech recognition model. 

For both speech recognition and regression model we used 70\% of the total data as training set, 10\% as validation set and remaining 20\% as test set. The Figure 3 shows the training and validation loss of the regression model.

\begin{figure}[h]
\begin{center}
\includegraphics[height=7cm,width=0.3\textwidth,trim={0.1cm 0.1cm 0.1cm 0.1cm},clip]{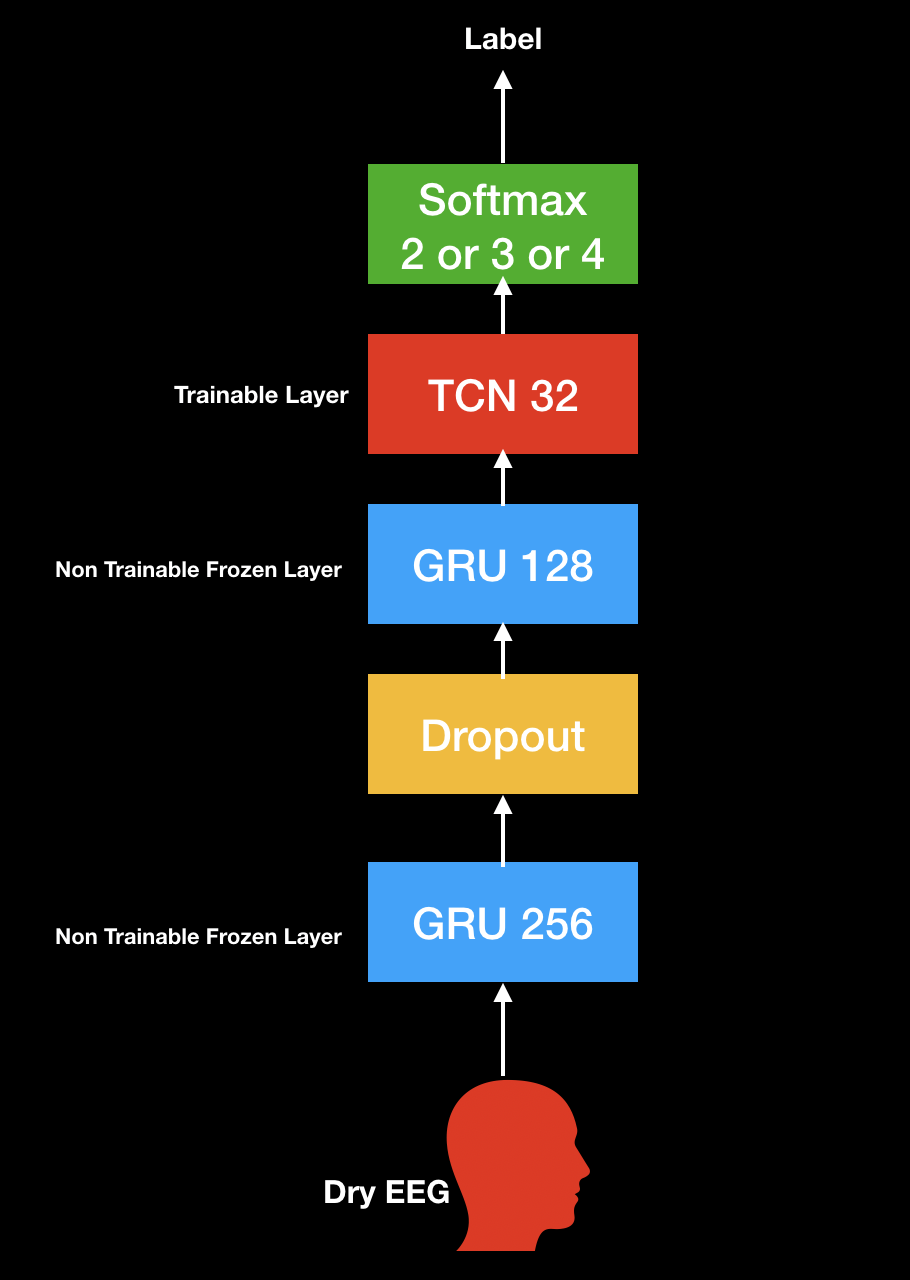}
\caption{Speech Recognition Model} 
\label{1vsall}
\end{center}
\end{figure}

\begin{figure}[h]
\begin{center}
\includegraphics[height=7cm,width=0.3\textwidth,trim={0.1cm 0.1cm 0.1cm 0.1cm},clip]{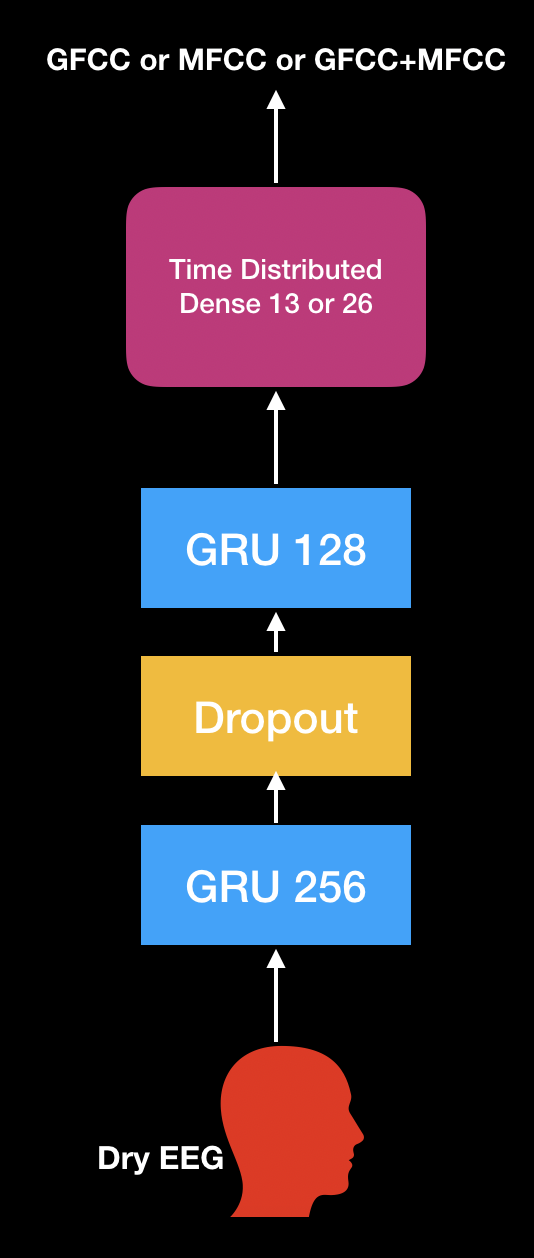}
\caption{Regression Model} 
\label{1vsall}
\end{center}
\end{figure}

\begin{figure}[h]
\begin{center}
\includegraphics[height=5cm,width=0.3\textwidth,trim={0.1cm 0.1cm 0.1cm 0.1cm},clip]{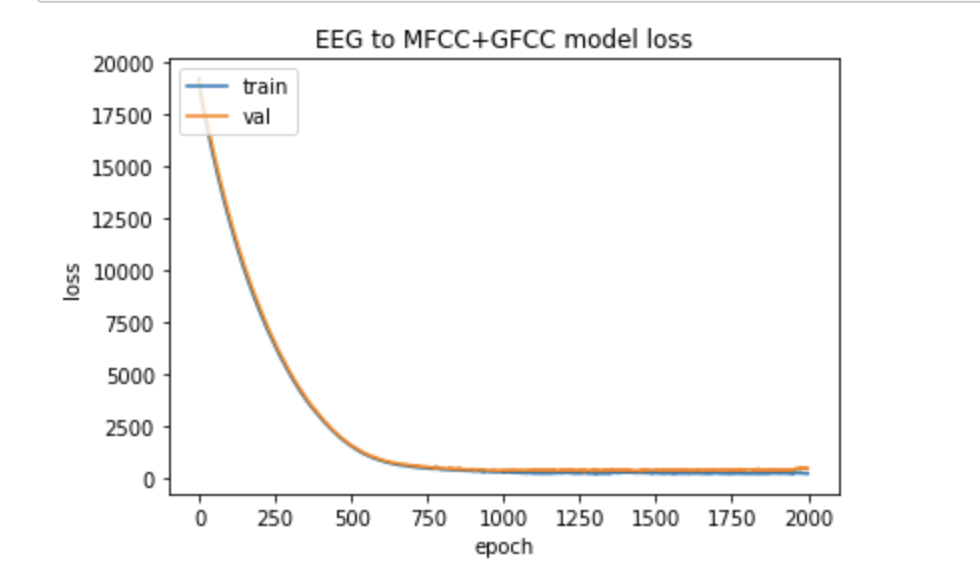}
\caption{Training and Validation loss for model used to predict MFCC and GFCC from EEG features} 
\label{1vsall}
\end{center}
\end{figure}

\section{Design of experiments for building the data set}
\label{sec:typestyle}

Two male and two female subjects took part in the Speech-EEG recording experiments to collect data. Out of the four subjects, three were native American English speakers. All the four subjects were UT Austin undergraduate students in their early twenties.
Each subject was asked to speak English vowels a,e,i and English word left and their simultaneous speech and EEG signals were recorded. The number of times the subjects were asked to repeat the experiment and number of samples recorded were similar to the study protocol explained in \cite{krishna2019speech}. The data was recorded in absence of background noise. 

We used QUICK-20 DRY EEG HEADSET for performing EEG recording experiments. The headset was manufactured by Cognionics. We used the data from the following 17 EEG sensors from the headset fp1, fp2, f8, f3, fz, f4,
c3, cz, p8, p7, pz, t3
,p3, o1, o2, c4 and t4. The sensors were placed on the headset following the standard international 10-20 layout.

\section{EEG feature extraction details}
\label{sec:majhead}

The EEG signals were sampled at 1000Hz and a fourth order IIR band pass filter with cut off frequencies 0.1Hz and 70Hz was applied. A notch filter with cut off frequency 60 Hz was used to remove the power line noise.
The EEGlab's \cite{delorme2004eeglab} Independent component analysis (ICA) toolbox was used to remove other biological signal artifacts like electrocardiography (ECG), electromyography (EMG), electrooculography (EOG) etc from the EEG signals.

Then we extracted five statistical features for EEG, namely root mean square, zero crossing rate, moving window average, kurtosis and power spectral entropy \cite{krishna2019speech,krishna20}.  In total there were 85 features (17(channels) X 5) for EEG signals. The EEG features were extracted at a sampling frequency of 500Hz for each EEG channel.

Even though in \cite{krishna20,krishna2019speech} authors performed non-linear dimension reduction after extracting EEG features, for this work we didn't perform dimension reduction as we were working with fewer number of EEG channels compared to the number of sensors used by authors in \cite{krishna20,krishna2019speech}.

\section{Acoustic Feature Extraction}
\label{sec:print}

For training the regression model we extracted MFCC and GFCC features each of dimension 13 from the recorded speech signal as acoustic features. The recorded speech signal was sampled at a sampling frequency of 16KHz.  The acoustic features were also extracted at the same sampling frequency of 500Hz like that of EEG features to avoid sequence-sequence mismatch.

\section{Results}
\label{sec:page}
We used test accuracy as the performance metric of the speech recognition model during test time. Test accuracy is defined as the ratio of number of correct predictions given by the model to total number of predictions in the test set.  The obtained test time results are summarized in Table 1. The results demonstrate that when the GRU layers of the speech recognition model were initialized with weights derived from EEG to MFCC + GFCC regression model, resulted in highest test time accuracy for all the experiments. We observed highest test accuracy value of \textbf{79.07 \%} for the experiment involving making predictions over first two labels as seen from Table 1. The Figure 4 shows the corresponding training and validation accuracy plot for the experiment. As seen from the plot we can observe that training and validation accuracy values were almost comparable, indicating our model didn't over fit on the data.  
 
When we performed speech recognition experiment using EEG features from frontal and temporal lobe sensors (total 8 channels) over the first two labels where the GRU layer weights were initialized with MFCC + GFCC regression weights, we observed a test accuracy of \textbf{69.77\%}. 
 
Our results were poor compared to isolated speech recognition using wet EEG demonstrated by authors in \cite{krishna2019speech} where they were able to achieve average test accuracy of more than 90\%, indicating wet EEG offer better SNR than dry EEG for the task of speech recognition even though it is more convenient for a subject to wear a wireless dry EEG headset compared to a wired wet EEG cap used by authors in \cite{krishna2019speech}.

\begin{table}[!ht]
\centering
\begin{tabular}{|l|l|l|l|l|}
\hline
\textbf{\begin{tabular}[c]{@{}l@{}}Number\\ of\\ Labels\\ Used\end{tabular}} & \textbf{\begin{tabular}[c]{@{}l@{}}\% Test\\ Accuracy\\ Baseline\\ (GRU\\  layers\\ Random\\  Weights)\end{tabular}} & \multicolumn{1}{c|}{\textbf{\begin{tabular}[c]{@{}c@{}}\% Test\\ Accuracy\\ (GRU layers\\ MFCC\\ Weights)\end{tabular}}} & \textbf{\begin{tabular}[c]{@{}l@{}}\% Test\\ Accuracy\\ (GRU layers\\ GFCC\\  Weights)\end{tabular}} & \textbf{\begin{tabular}[c]{@{}l@{}}\% Test\\ Accuracy\\ (GRU layers\\ MFCC \\ +\\ GFCC\\ Weights)\end{tabular}} \\ \hline
2                                                                            & 74.42                                                                                                                & 77.91                                                                                                                    & 73.26                                                                                                & \textbf{79.07}                                                                                                  \\ \hline
3                                                                            & 62.90                                                                                                                & 70.16                                                                                                                    & 70.16                                                                                                & \textbf{72.58}                                                                                                  \\ \hline
4                                                                            & 50.65                                                                                                                & 56.49                                                                                                                    & 55.19                                                                                                & \textbf{61.04}                                                                                                  \\ \hline
\end{tabular}
\caption{Speech Recognition Test Time Results}
\end{table}

\begin{figure}[h]
\begin{center}
\includegraphics[height=6.5cm,width=0.5\textwidth,trim={0.1cm 0.1cm 0.1cm 0.1cm},clip]{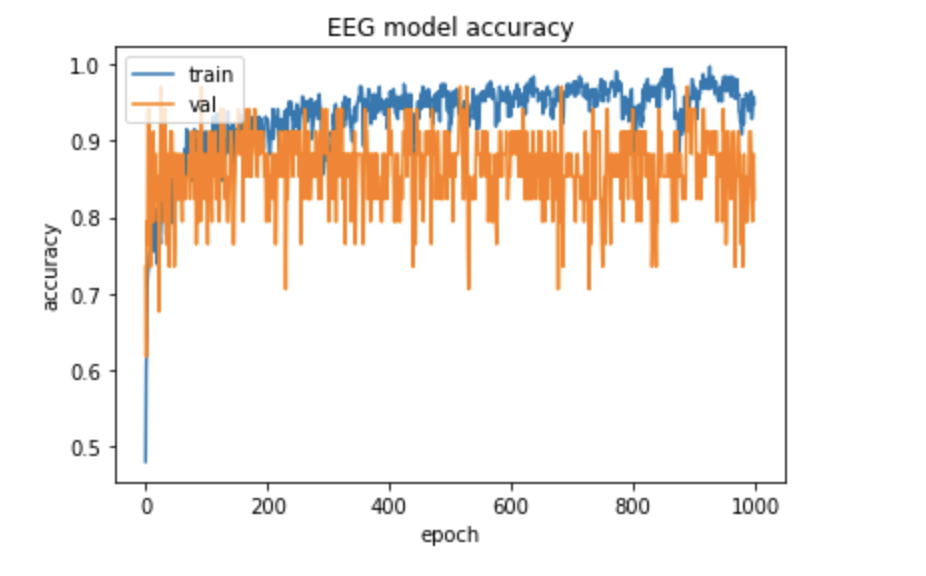}
\caption{Training and Validation accuracy plot} 
\label{1vsall}
\end{center}
\end{figure}

\section{Conclusions and Future work}
\label{sec:illust}
In this paper we demonstrated the feasibility of using dry EEG features for performing isolated speech recognition on a limited English vocabulary using deep learning model. To the best of our knowledge this is the first work which explored speech recognition using dry EEG features. We were able to achieve highest test accuracy of \textbf{79.07\%}. 

Future work will focus on exploring the use of dry EEG for other speech technologies and improving current performance.

\section{Acknowledgements}
\label{sec:foot}
We would like to thank Kerry Loader and Rezwanul Kabir from Dell, Austin, TX for donating us the GPU to train the models used in this work.




\bibliographystyle{IEEEbib}
\bibliography{strings,refs}

\end{document}